\documentclass{WileyMSP-template}
\usepackage{xcolor}
\usepackage{graphicx}
\usepackage{amsmath}
\usepackage{amssymb}
\usepackage{mathtools}
\usepackage{amsthm}
\usepackage{physics}
\usepackage{gensymb}
\usepackage{color}
\usepackage{url}
\usepackage{tabularx}
\usepackage{soul}

\begin{document}

\pagestyle{fancy}
\rhead{}

\title{Digitized Phase Change Material Heterostack for Diffractive Optical Neural Network}

\maketitle

% Author: Please give full first and last names for authors and include * after the name of all corresponding authors

\author{Ruiyang Chen}
\author{Cunxi Yu}
\author{Weilu Gao*}

% Dedication
% \dedication{Optional dedication here. If no dedication is required, please leave blank}

\begin{affiliations}
Ruiyang Chen, Weilu Gao\\
Department of Electrical and Computer Engineering, The University of Utah, Salt Lake City, UT 84112, USA\\
Cunxi Yu\\
The University of Maryland, College Park, Department of Electrical and Computer Engineering, College Park, Maryland 20742, USA\\
$*$To whom correspondence should be addressed; Email Address: weilu.gao@utah.edu
\end{affiliations}

% Keywords: Please provide a minimum of three and a maximum of seven keywords, separated by commas


% Abstract should be written in the present tense and impersonal style (i.e., avoid we), and be at most 200 words long
\begin{abstract}
All-optical and fully reconfigurable diffractive optical neural network (DONN) architectures are promising for high-throughput and energy-efficient machine learning (ML) hardware accelerators for broad applications. However, current device and system implementations have limited performance. This work demonstrates a novel diffractive device architecture, which is named digitized heterostack and consists of multiple layers of nonvolatile phase change materials (PCMs) with different thicknesses. This architecture can both leverage the advantages of PCM optical properties and mitigate challenges associated with implementing multilevel operations in a single PCM layer. Proof-of-concept experiments demonstrate the electrical tuning of one PCM layer in a spatial light modulation device, and thermal analysis guides the design of DONN devices and systems to avoid thermal crosstalk if individual heterostacks are assembled into an array. Further, heterostacks containing three PCM layers are designed to have a large phase modulation range and uniform coverage and the ML performance of DONN systems with designed heterostacks is evaluated. The developed device architecture provides new opportunities for desirable energy-efficient, fast-reconfigured, and compact DONN systems in the future.   
\end{abstract}

% Text: Please use section headings and subheadings as specified below. For communications, all section headings apart from Experimental Section should be removed
% Please make the first reference to a display item bold: \textbf{Figure 1}
% Do not abbreviate Figure, Equation, etc.; display items are always singular, i.e., Figure 1 and 2.
% Equations are always singular, i.e., Equation 1 and 2, and should be inserted using the {equation} environment, not as graphics
% Please do not use footnotes in the text, additional information can be added to the Reference list.

\section{Introduction}

Machine learning (ML) has transformed a broad range of applications, including imaging and sensing~\cite{LeCunEtAl2015N,GoodfellowEtAl2016,RodriguesEtAl2021NP}, chip design~\cite{MirhoseiniEtAl2021N}, and scientific discovery~\cite{ButlerEtAl2018N,SeniorEtAl2020N}. With the explosively increasing size of ML models, such as ChatGPT, their execution on hardware requires unsustainably large computational resources and energy consumption. Alternative to electronic platforms, optical computing platforms have recently gained much interest as new high-throughput and energy-efficient hardware ML accelerators thanks to the extreme parallelism and low static energy consumption of photons~\cite{WetzsteinEtAl2020N}. For example, two-dimensional (2D) integrated photonics-based processors with different types of on-chip optoelectronic  modulators can perform various ML tasks and mathematical operations~\cite{ShenEtAl2017NP,RiosEtAl2019SA,FeldmannEtAl2021N,FengEtAl2024N}. In addition, three-dimensional (3D) free-space optical systems~\cite{LinEtAl2018S,HamerlyEtAl2019PR,MennelEtAl2020N,SpallEtAl2020OL,MiscuglioEtAl2020O,ZhouEtAl2021NP,GaoEtAl2021APR,WangEtAl2022NC,ChenEtAl2022LPR,ChenEtAl2023AIS,FanEtAl2023AIS,HuEtAl2024NC}, which spatially manipulate the amplitude, phase, and polarization of light with engineered 2D surfaces, advance optical computing capabilities. In particular, diffractive optical neural network (DONN) systems operating based on the physics of spatial light modulation and optical diffraction enable ML functionalities~\cite{HuEtAl2024NC}. 

\medskip
Current DONN systems are mainly passive, meaning diffractive arrays cannot be on-demand reconfigured for different ML tasks once they are fabricated. Note that the reconfiguration of diffractive arrays occurs only when ML tasks change, which is not frequent and is distinct from encoding data streams using fast electro-optic modulators.  Limited reconfigurability was implemented in a hybrid optoelectronic DONN system~\cite{ZhouEtAl2021NP}, where electrical-to-optical (E/O) and optical-to-electrical (O/E) conversions occurred between diffractive arrays. However, these intermediate conversions increased energy consumption and processing latency, and the induced deployment errors required additional iterative adaptive tuning processes. We recently made a major advance toward this ultimate goal by constructing a fully reconfigurable DONN system based on cascaded liquid crystal spatial light modulators in the visible wavelength range~\cite{ChenEtAl2022LPR}. This system is all-optical without E/O and O/E conversions. However, external voltage and current are needed to maintain the rotation states of liquid crystals when performing the inference of a specific ML task. The reconfiguration speed is limited to hundreds of milliseconds. The lateral size of each liquid crystal pixel is also much larger than the operation wavelength. The further improvement toward the most desirable diffractive arrays in DONN systems, which are energy-efficient without maintaining external stimulus, fast reconfigured, and of compact footprint, relies on new materials and device architectures. 

\medskip
Here, we present a novel near-infrared diffractive device based on nonvolatile chalcogenide phase change materials (PCMs), which can be beneficial for DONN systems from multiple perspectives. First, the material states of PCMs can be fast and reversibly reconfigured between crystalline or amorphous phases with an electrical or optical pulse~\cite{WuttigEtAl2017NP,AbdollahramezaniEtAl2020N,ZhangEtAl2021APL}. Second, the material states can be preserved after reconfiguration without external stimulus for $>10$ years~\cite{WuttigEtAl2007NM,WongEtAl2010PI} such that static energy consumption is minimal~\cite{SebastianEtAl2020NN}. Third, the change of optical refractive indices can be $>1$ over a wide spectral range~\cite{WuttigEtAl2017NP}, enabling compact and broadband operations for integrated~\cite{RiosEtAl2015NP,ChengEtAl2017SA,RiosEtAl2019SA,ZhangEtAl2019NC,FangEtAl2022NN} and free-space photonic components and systems~\cite{WangEtAl2016NP,WangEtAl2021NN,ZhangEtAl2021NN,AbdollahramezaniEtAl2022NC}. Fourth, PCMs are scalable~\cite{RaouxEtAl2008JRD} and compatible with other materials for constructing complex architectures. The capability of handling complex ML tasks in DONN systems is naturally dependent on the number of modulation levels achieved in each diffractive pixel. Multilevel operations in practical PCM-based devices are achieved with multiple intermediate states, which are a mixture of crystalline and amorphous phases and typically implemented by carefully designing electrical pulse profiles to deliver various heating energy~\cite{RiosEtAl2015NP,ZhangEtAl2019P,LiEtAl2019O,ZhangEtAl2021NN,AbdollahramezaniEtAl2022NC}. However, uniformly and reliably achieving many levels (e.g., $>10$) in a single PCM layer is still challenging and the formation of intermediate states is intrinsically stochastic~\cite{WangEtAl2021CM2}, which can potentially lead to nonuniform and unreliable operations in large-area devices. Hence, to mitigate these challenges associated with multilevel operations in a single PCM layer, we demonstrate a digitized PCM heterostack architecture containing multiple PCM films with different thicknesses and other dielectrics. Each PCM film is only under crystalline and amorphous phases or with only a few intermediate states. We fabricated and experimentally characterized a proof-of-concept device for spatial light modulation, and performed thermal analysis to determine the needed spacing between PCM films in the heterostack and spacing between heterostack pixels to avoid thermal crosstalk if individual diffractive devices are assembled into an array for the DONN system. Together with the obtained proof-of-concept experimental results, we further designed layer thicknesses in a heterostack containing three PCM layers to have a large phase modulation range and uniform coverage and finally evaluated the performance of DONN systems with designed heterostacks. 

% \subsection{First Subsection}
% \subsubsection{First Sub Subsection}
% \threesubsection{First lowest-level subsection}

\section{Results}

\begin{figure}
    \centering
    \includegraphics[width=0.6\columnwidth]{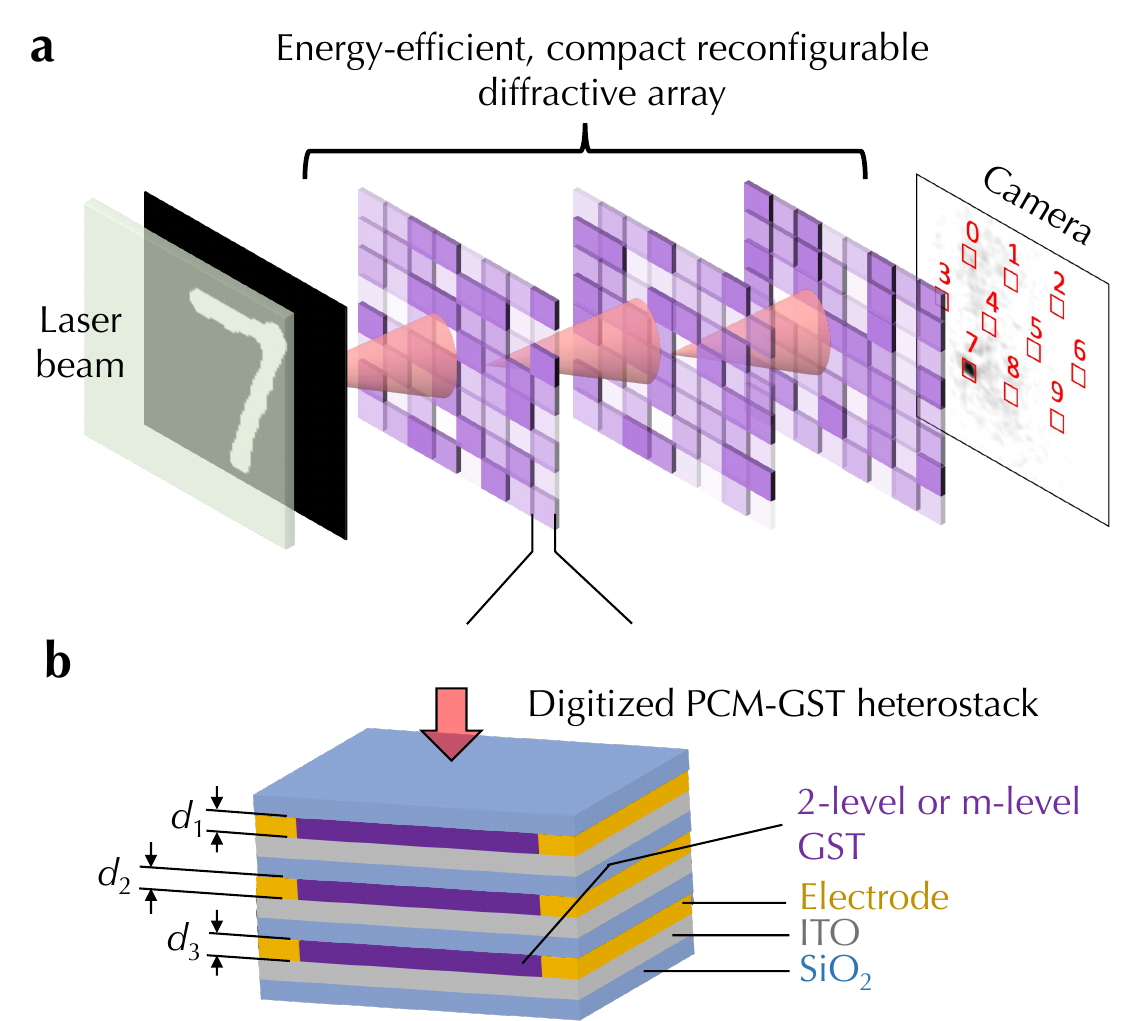}
    \caption{Illustrations of (a)~a DONN system with cascaded energy-efficient, compact reconfigurable diffractive arrays with each pixel implemented using a (b)~digitized GST heterostack.}
    \label{fig:fig1}
\end{figure}

Figure\,\ref{fig:fig1}a illustrates the general architecture of a DONN system consisting of multiple cascaded diffractive arrays. The input laser beam is incident on images, diffracted by cascaded diffractive arrays, and captured on a camera that is essentially a 2D photodetector array. On the camera, there are multiple pre-defined localized regions to represent the meaning of input images. For example, we can define 10 regions to represent 10 handwritten digits in the Modified National Institute of Standards and Technology (MNIST) database. Each pixel on a diffractive array can modulate the amplitude and phase of transmitted light. By training (i.e., optimizing) excitation and thus responses of each pixel on all diffractive arrays, the light wavefront after diffraction can converge into a pre-defined localized region and hence the DONN system can perform classification ML tasks of input images. 

\medskip
To create an energy-efficient, compact reconfigurable diffractive array, we developed a digitized PCM heterostack for each pixel as illustrated in Fig.\,\ref{fig:fig1}b. The specific PCM we utilized is the widely demonstrated Ge$_2$Se$_2$Te$_5$ (GST) material. There are multiple stacks of GST films with different thicknesses, such as three thicknesses \{$d_1, d_2, d_3$\} in Fig.\,\ref{fig:fig1}b. Each GST layer has an independent heater and electrodes to control GST phases in fully crystalline or amorphous phases, which is called a 2-level GST operation. Since layer thicknesses are different, the amplitude and phase modulation from the phase transition in each GST layer have different weights in total modulation of transmitted light, which is the principle of implementing multilevel operations. This is analogous to the conversion from binary to decimal numeral systems. For example, each 1 in a binary number (111)$_2$ has a different weight in its corresponding decimal number. The leading 1 represents 4, the second 1 represents 2, and the trailing 1 represents 1. The weight of each 1 is similar to the thickness of each GST layer, and 0 (1) on each digit is similar to the fully amorphous (crystalline) phase. Because of the existence of multiple reflections between layers, the thicknesses $d_1$, $d_2$, and $d_3$ in the 3-layer architecture may not follow the ratio $1:2:4$ as in binary numbers. Instead, these thicknesses can be optimized freely to have a large range and uniformly distributed multilevel modulation. Further, the 2-level GST operation can be extended to a few levels (e.g., 5 levels) with some relatively stable and easy-to-achieve intermediate states, named $m$-level operation. Practically, the as-fabricated heterostack with as-deposited amorphous GST films is first globally heated through hotplates or ovens to have crystalline GST films as the initial state, and then each GST layer is reconfigured sequentially using electrical pulses. The heterostack can be reset to the initial state through global heating or large electrical pulses. 

\begin{figure}
    \centering
    \includegraphics[width=0.7\columnwidth]{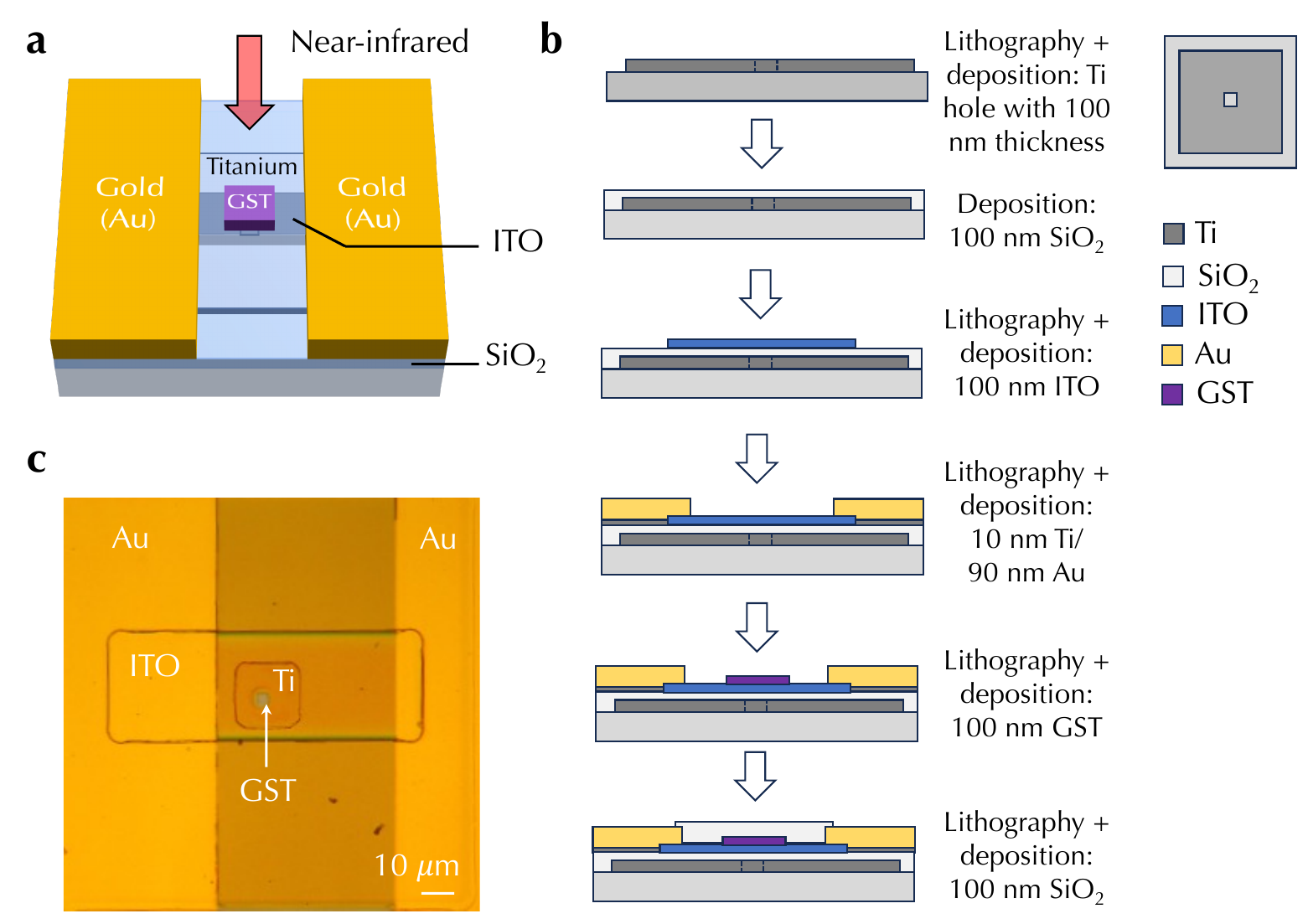}
    \caption{(a)~Illustration of the spatial light modulation device demonstrating the electrical tuning of one GST film. (b)~Flow chart of nanofabrication process. (c)~Photo of a manufactured device.}
    \label{fig:fig2}
\end{figure}

\medskip
For a proof-of-concept demonstration of the electrical tuning of one GST film in a spatial light modulation device, as illustrated in Fig.\,\ref{fig:fig2}a, we experimentally fabricated the device containing one nanopatterned GST film that can be electrically driven by an ITO-based heater through gold electrodes. The operation wavelength was chosen in the near-infrared range because of the relatively low loss of GST and ITO. The substrate is glass, which is also transparent at this wavelength. Figure\,\ref{fig:fig2}b displays the flowchart of nanofabrication processes, including multiple steps of lithography, deposition, and lift-off. More details can be found in \emph{Methods}. Fig.\,\ref{fig:fig2}c shows the photo of a manufactured device. 

\begin{figure}
    \centering
    \includegraphics[width=\columnwidth]{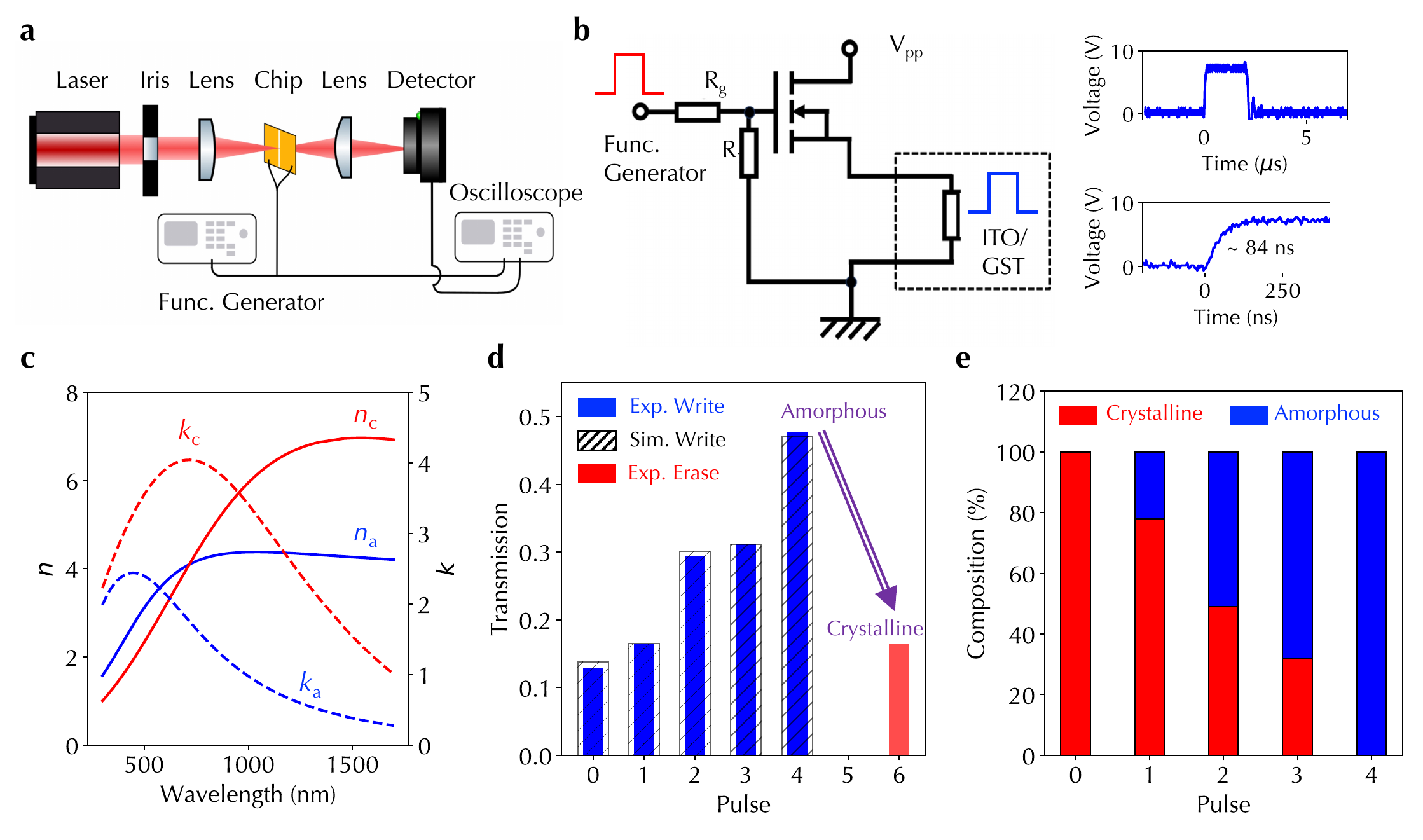}
    \caption{(a)~Illustration of the laser characterization setup. (b)~Diagram of the electrical driving circuit and corresponding driving voltage waveforms. (c)~Experimentally measured refractive index ($n$) and extinction coefficient ($k$) of the GST film. (d)~Experimentally measured and simulated transmittance under multiple writing and erasing pulses. (e)~Composition ratios of crystalline and amorphous phases under different pulses.}
    \label{fig:fig3}
\end{figure}

\medskip
Figure\,\ref{fig:fig3}a illustrates the optoelectronic characterization setup to characterize manufactured devices. Specifically, we focused a 1550\,nm laser onto the manufactured chip using lenses and the focused spot size was $\sim90\,\mu\mathrm{m}\times90\,\mu$m. Because of the relatively large laser spot size and small GST area ($\sim30\,\mu\mathrm{m}\times30\,\mu$m), we created a metallic titanium hole ($\sim10\,\mu\mathrm{m}\times10\,\mu$m) to block all other light so that the transmitted light passing through the device is from the GST film; see Fig.\,\ref{fig:fig2}b and Fig.\,\ref{fig:fig2}c as well. A function generator was used to drive a circuit to apply electrical pulses on the ITO heater and the oscilloscope was used to measure the time-dependent response from the driving circuit and a fast InGaAs detector. Figure\,\ref{fig:fig3}b displays the diagram of the driving electrical circuit. A fixed square voltage wave from the function generator was the input to the gate of a power transistor, whose source was connected to a DC voltage source with a controllable voltage $V_\mathrm{pp}$. More details can be found in \emph{Methods}. Hence, there was a voltage pulse applied to the ITO heater to tune the phase of GST film (Fig.\,\ref{fig:fig3}b). The obtained pulse width on the ITO heater was determined by the input square wave and the circuit response, and the pulse height was determined by $V_\mathrm{pp}$. By tuning $V_\mathrm{pp}$, we changed the pulse height. Figure\,\ref{fig:fig3}b also displays time-domain waveforms of one representative applied voltage pulse. The pulse width, rising time, and peak voltage were $\sim2\,\mu$s, $\sim84\,$ns, and $\sim7.5$V, respectively. 

\medskip
The as-deposited GST film was in the amorphous phase and then heated up on a hot plate to be converted to the fully crystalline phase. We measured the refractive index ($n$) and extinction coefficient ($k$) in a wavelength range of $300 - 1700$\,nm using ellipsometry; see Fig.\,\ref{fig:fig3}c and more details in \emph{Methods}. We set $V_\mathrm{pp}$ at 19\,V and applied multiple pulses to drive devices and switch the GST film from the crystalline phase to the amorphous phase. Blue bars in Fig.\,\ref{fig:fig3}d display the measured transmittance, which is defined as the ratio of the transmitted power passing through the GST film over the incident power passing through other layers except the GST film, after several writing pulses. ``0'' pulse means the initial state, which is the crystalline phase. Note that the measured transmittance was calibrated based on the laser spot and Ti pinhole sizes. We observed a 5-level GST operation. Further, we applied two additional pulses as erasing pulses to switch the amorphous phase back to the crystalline phase. Moreover, with experimentally obtained $n$ and $k$ of the GST film, available optical constants of other materials~\cite{TangEtAl2022LPR}, and experimental film thicknesses in the stack, we employed a standard $2\times2$ transfer matrix method~\cite{YehEtAl2005} to simulate the transmittance of the stack at 1550 nm wavelength. For intermediate states, we assumed a linear combination of crystalline and amorphous phases and corresponding $n$ and $k$. Specifically, the composition ratio of the crystalline phase was denoted as $\alpha_\mathrm{c}$ and the amorphous ratio $\alpha_\mathrm{a}$ is $1 - \alpha_\mathrm{c}$. Hence, $n$ and $k$ for intermediate states are $\alpha_\mathrm{c}n_\mathrm{c} + \alpha_\mathrm{a}n_\mathrm{a}$ and $\alpha_\mathrm{c}k_\mathrm{c} + \alpha_\mathrm{a}k_\mathrm{a}$, respectively, where $n_\mathrm{c}$ ($n_\mathrm{a}$) and $k_\mathrm{c}$ ($k_\mathrm{a}$) are $n$ and $k$ for the crystalline (amorphous) phase shown in Fig.\,\ref{fig:fig3}c. $\alpha_\mathrm{c}$ was a fitting parameter during simulations. Black-striped bars in Fig.\,\ref{fig:fig3}d show simulation results under writing pulses, agreeing well with experimental results. Figure\,\ref{fig:fig3}e displays the composition ratio of amorphous and crystalline phases at intermediate states and we observed a gradual and nearly linear phase change under writing pulses. 

\begin{figure}
    \centering
    \includegraphics[width=0.8\columnwidth]{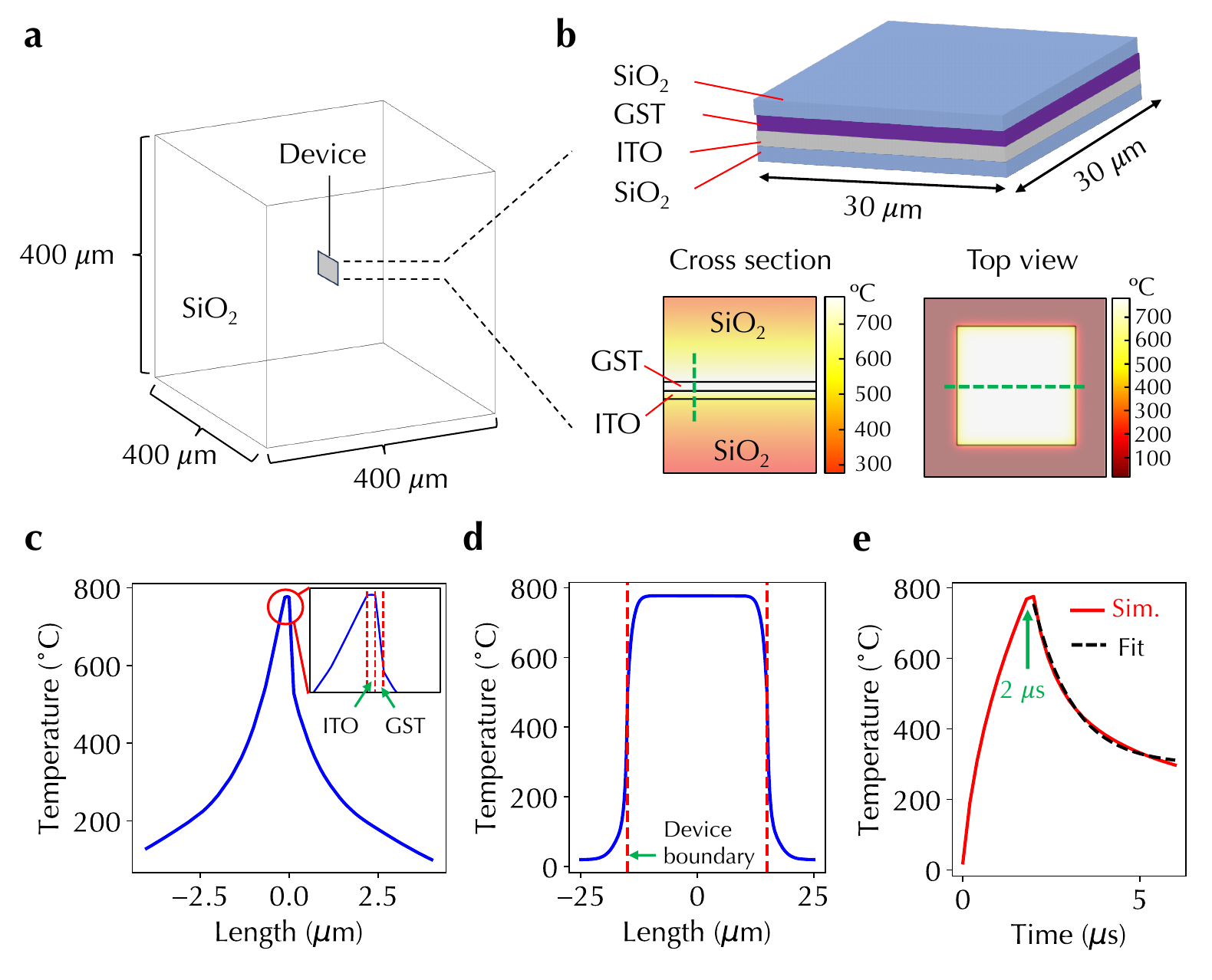}
    \caption{(a)~Schematic of COMSOL simulations. (b)~Illustration of a heterostack with one GST film, and the cross section and top view of temperature distribution under experimental pulses. (c)~Temperature profile across the green line in the cross-section of temperature distribution. (d)~Temperature profile across the green in the top view of temperature distribution. (e)~Time-dependent response of the highest temperature in the GST film.}
    \label{fig:fig4}
\end{figure}

\medskip
In the digitized heterostack shown in Fig.\,\ref{fig:fig1}b, the SiO$_2$ spacing layers between GST layers are necessary to minimize the thermal crosstalk that can induce the phase transition from amorphous to crystalline phases and have independent heater control for each GST layer. In addition, when heterostack pixels are grouped into a 2D diffractive array (Fig.\,\ref{fig:fig1}a) for a DONN system, the lateral spacing between pixels is also needed for minimal crosstalk between pixels. To analyze the temperature profiles of the GST film under experimental electrical pulses, we performed COMSOL Multiphysics simulations. 
% The purpose of such simulations is to understand the temperature profile of the GST film in a realistic digitalized GST heterostack for DONNs so that the diffractive unit pixel can be designed to avoid thermal crosstalks. 
Figure\,\ref{fig:fig4}a displays the diagram and dimensions of COMSOL simulations. The device was at the center of the simulation region, which was filled with SiO$_2$. The top panel of Fig.\,\ref{fig:fig4}b shows the schematic of the simulated device consisting of one GST layer and one ITO heater. The size was set as $30\,\mu\mathrm{m}\times30\,\mu$m. More details can be found in \emph{Methods}. The bottom panels of Fig.\,\ref{fig:fig4}c display the cross-section and top view of temperature distribution in the device under experimental writing pulses. Figure\,\ref{fig:fig4}c shows the temperature profile across the green line in the cross-section view in Fig.\,\ref{fig:fig4}b and the highest temperature achieved in the GST layer is $\sim780^{\circ}$C. This temperature is sufficient to induce the phase transition from crystalline to amorphous phases, typically requiring the melting temperature $>600^{\circ}$C. The one-side temperature decay length from the highest temperature to a temperature close to $\sim100^{\circ}$C is $\sim4\,\mu$m and the transition temperature from amorphous to crystalline phases is $>150^{\circ}$C, meaning that within the heterostack the thickness of SiO$_2$ spacing layers needs to $\geq4\,\mu$m to minimize the thermal crosstalk. Similarly, Fig.\,\ref{fig:fig4}d shows the temperature profile across the green line in the top view in Fig.\,\ref{fig:fig4}b. At a distance $5\,\mu$m away from the device boundary, the temperature becomes close to the room temperature ($\sim30^{\circ}$C). Hence, in lateral dimensions, the pixel spacing $\geq5\,\mu$m can avoid the thermal crosstalk. Figure\,\ref{fig:fig4}e shows the time-dependent response of the highest temperature observed in Fig.\,\ref{fig:fig4}c and Fig.\,\ref{fig:fig4}d. The peak temperature occurs at the end of pulse $2\,\mu$s. After the excitation, the temperature drops due to the heat dissipation in the surrounding environment. We fit the temperature decay using an exponential function and obtained a time constant $\sim1.12\,\mu$s. 

\begin{figure}
    \centering
    \includegraphics[width=\columnwidth]{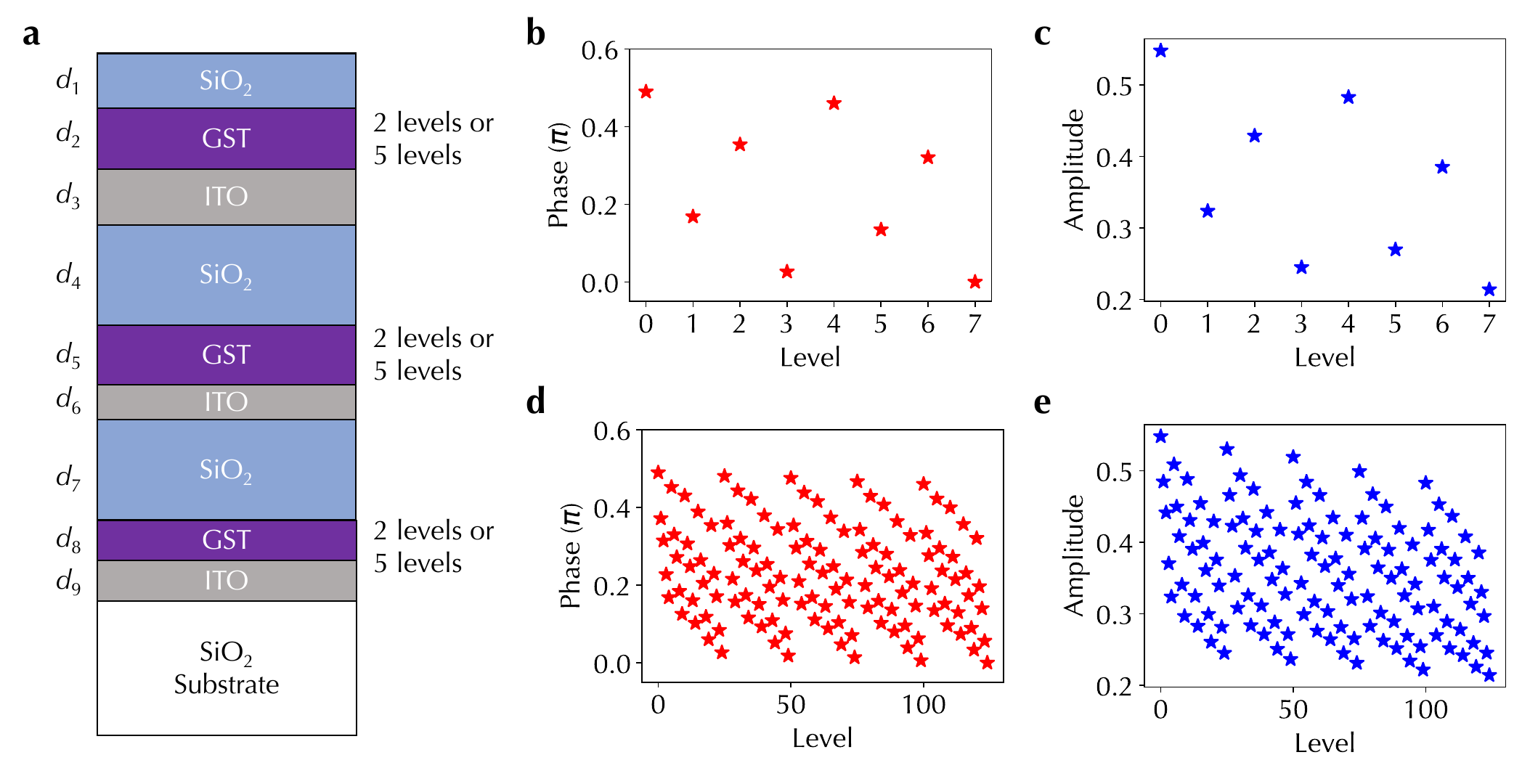}
    \caption{(a)~Diagram of a digitized GST heterostack containing 3 GST layers with each layer operating under 2-level or 5-level operations. (b)~Phase and (c)~amplitude response of transmitted light under different combinations of levels when the GST film can achieve 2 states. (d)~Phase and (e)~amplitude response of transmitted light under different combinations of levels when the GST film can achieve 5 states.}
    \label{fig:fig5}
\end{figure}

\medskip
With experimentally measured transmission properties of one GST film, we further designed thicknesses of SiO$_2$, ITO, and GST films in a heterostack containing three layers of GST films using the transfer matrix method, as illustrated in Fig.\,\ref{fig:fig5}a. All thicknesses \{$d_1, d_2, ..., d_{9}$\} were designed to have large and uniform phase responses. Each GST film can perform either under 2-level operation with only crystalline and amorphous phases or under 5-level operation with intermediate states with the experimentally obtained composition ratios shown in Fig.\,\ref{fig:fig3}e. The thickness of SiO$_2$ capping layer ($d_1$) was fixed as 100\,nm and the thicknesses of SiO$_2$ spacing layer ($d_4$ and $d_7$) were required to be $\geq 4\mu$m to avoid thermal crosstalks based on the COMSOL simulations in Fig.\,\ref{fig:fig4}. The optimized ITO thickness was 122\,nm and the thicknesses of three GST layers ($d_2, d_5$, and $d_8$) were 10\,nm, 60\,nm, and 80\,nm, respectively. Figure\,\ref{fig:fig5}b and Fig.\,\ref{fig:fig5}c display the phase and amplitude response of transmitted light under different combinations of levels when the GST film was under the 2-level operation. There are in total $8=2^3$ different combinations for 3 GST films. The phase modulation range is $\sim0.5\pi$. Further, Fig.\,\ref{fig:fig5}d and Fig.\,\ref{fig:fig5}e display the phase and amplitude response of transmitted light under different combinations of levels when the GST film was under the 5-level operation. In total, there are $125 = 5^3$ combinations for three GST films and hence there are more intermediate phase and amplitude response points in 5-level-operation GST films compared to 2-level-operation GST films. 

\begin{figure}
    \centering
    \includegraphics[width=0.8\columnwidth]{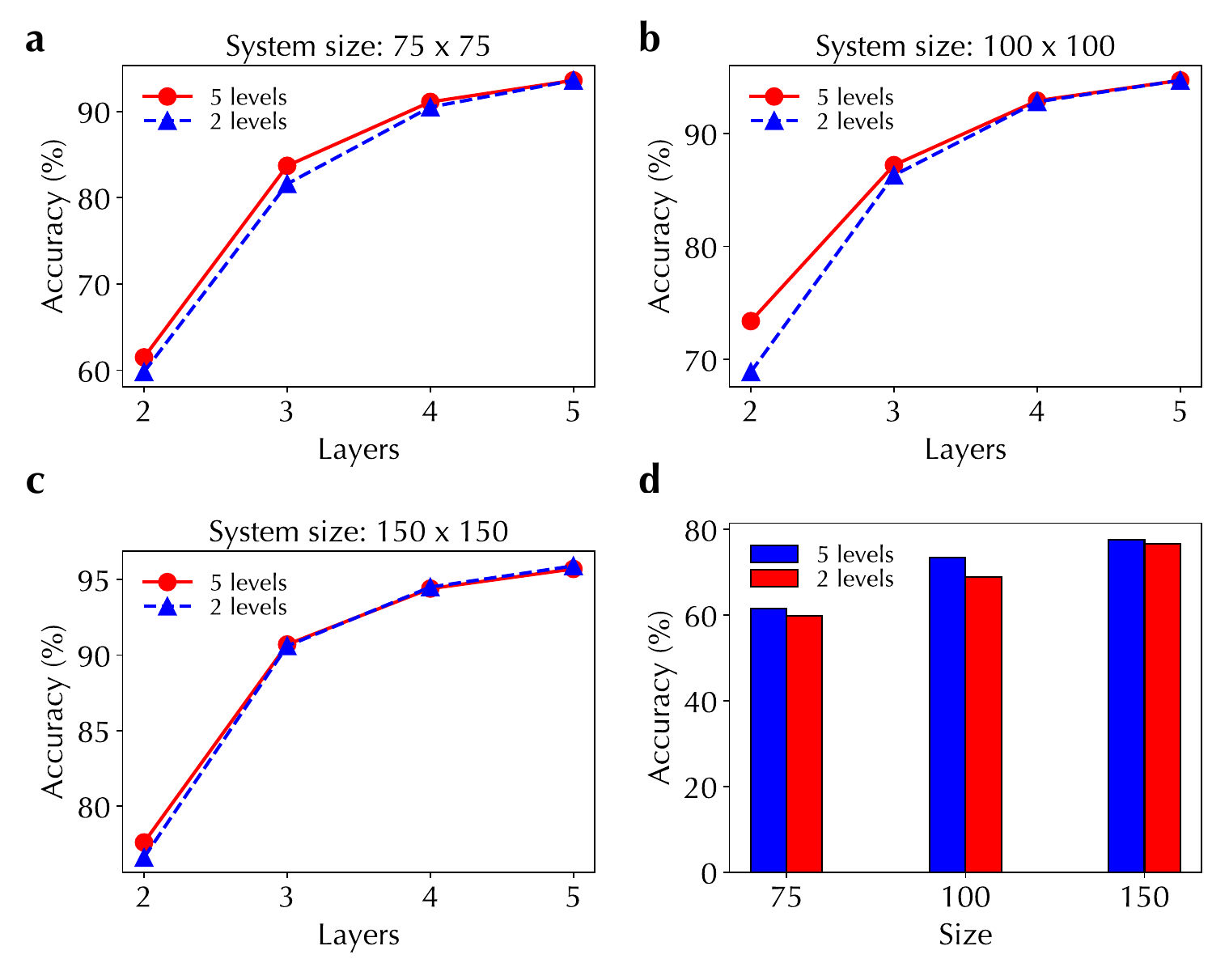}
    \caption{(a)--(c)~Classification accuracies of DONNs of different system sizes and layers with digitalized GST heterostack operating under 2 or 5 levels for each GST film. (d)~Comparison of different system sizes for 2-level and 5-level GST films in 2-layer DONN systems.}
    \label{fig:fig6}
\end{figure}

\medskip
Finally, we evaluated the performance of a DONN system for classifying handwritten digits from the MNIST dataset, if the individual heterostack diffractive device designed in Fig.\,\ref{fig:fig5} can be assembled into a 2D diffractive array. Although not experimentally demonstrated here, this type of 2D array is experimentally feasible. For example, a recent work has demonstrated a $8000\times2000$ GST array with a small pixel pitch $1.5\,\mu$m~\cite{KimEtAl2019J}. To train diffractive arrays, meaning finding the best combination of levels for all heterostack diffractive pixels, through the backpropagation algorithm in ML software frameworks, we need to incorporate the irregular amplitude and phase device responses into DONN calculations in a differentiable manner. However, the device responses cannot be fitted with analytical equations and we utilized a Gumbel-Softmax reparameterization approach to approximate discrete distributions with a differentiable continuous function~\cite{ChenEtAl2022LPR}. Each pixel size was set as $30\,\mu\mathrm{m}\times30\,\mu$m to be consistent with experiments in Fig.\,\ref{fig:fig2} and a $5\,\mu$m spacing between pixels was set to avoid thermal crosstalk. Hence, the pixel filling factor was 56.25\,\%. We evaluated DONN systems with different array sizes, number of layers, and heterostacks containing 2-level and 5-level GST films, as summarized in Fig.\,\ref{fig:fig6}. As expected, classification accuracies increase with increasing array size and layer number (Fig.\,\ref{fig:fig6}a -- c). Further, for 2-layer DONN systems, the heterostack with 5-level GST films improves the DONN system classification performance compared to 2-level GST films, as shown in Fig.\,\ref{fig:fig6}d. 

\section{Conclusion and Discussion}
We demonstrated a feasible design of a novel diffractive device for DONN systems based on a PCM heterostack, which can leverage the advantages of PCM optical properties and mitigate challenges associated with implementing multilevel operations.  We performed not only proof-of-concept experiments but also detailed analyses on both device and system levels. The cyclability of current devices is not good because of the poor quality of ITO heaters. More thermally stable, efficient, and transparent heaters, such as carbon nanomaterials including graphene~\cite{FangEtAl2022NN} and aligned carbon nanotube films~\cite{HeEtAl2016NN}, can substantially improve the efficiency and cyclability of heaters. Further, in addition to GST materials, the utilized PCM in heterostacks can be extended to other materials with lower loss for broadband operations, such as Sb$_2$Se$_3$~\cite{DelaneyEtAl2020AFM}. 

% Experimental section
\section{Methods}
\subsection{Device fabrication}
The overall nanofabrication flow is illustrated in Fig.\,\ref{fig:fig2}b. Specifically, a $10\,\mu\mathrm{m}\times10\,\mu$m hole was defined by photolithography using a mask aligner Suss MA1006 and a 100-nm thick Ti film was deposited using a Denton Discovery 18 sputtering system. The Ti hole structure was then formed after a lift-off process. A 100-nm thick SiO$_{2}$ film was then deposited on top using a Denton 635 sputtering system. A $150\,\mu\mathrm{m}\times50\,\mu$m area was defined by photolithography and a 100-nm thick ITO film was deposited using the Denton Discovery 18 system to form a heater after lift-off. The ITO heater was connected by two electrodes, which contained 10-nm thick Ti and 90-nm Au defined by photolithography and deposited by the Denton 635 system. Afterward, a $30\,\mu\mathrm{m}\times30\,\mu$m area was defined by photolithography and a 100-nm thick GST film was deposited using a Denton Discovery 18 system at an argon pressure of 4.5\,mTorr and a power setting of 35\,W. 

\subsection{Optoelectronic characterization setup} 
The schematic diagram of the experimental setup is illustrated in Figure\,\ref{fig:fig3}a. The employed laser diode was LQC1550-05E from Newport Corporation with a center wavelength $1550\,$nm. The elliptical output beam from the laser diode was reshaped to a round beam through lenses and an iris. Manufactured chips were driven by an electrical circuit connected to a function generator (Tektronix AFG2020). In the driving circuit, the model of the employed power transistor was Infineon IRLZ34N and the model of the employed DC source was Tektronix PS281. The output light power was measured by an InGaAs detector with a 5\,GHz speed bandwidth and an operation wavelength range of $800 - 1700$\,nm (DET08C from Thorlabs, Inc.). The time responses of driving electrical pulses and the detector reading were recorded by an oscilloscope with a bandwidth of 300\,MHz (Tektronix TDS3034).

\subsection{Ellipsometry measurement} 
The film thickness and optical refractive index $n$ and extinction coefficient $k$ of GST films were measured using a J.A. Woollam Variable Angle Spectroscopic Ellipsometer (VASE) over a wavelength range of $300-1700$\,nm. In the measurement, polarized light was reflected off the sample surface, and the change in polarization was measured as two quantities: amplitude ratio $\Psi$ and phase difference $\Delta$. By fitting measurements with a model describing materials and sample structures, the optimal film thickness and optical indices were obtained to minimize the error between experiments and calculations. 

\subsection{COMSOL simulations} 
A 3D finite element simulation using COMSOL Multiphysics was conducted to analyze the temperature distribution in Fig.\,\ref{fig:fig4}. The ITO thickness was 120\,nm and the GST thickness was 130\,nm. The thermal conductivity of ITO was set as 11\,W m$^{-1}$K$^{-1}$ and the GST thermal conductivity was set as 0.27\,W m$^{-1}$ K$^{-1}$~\cite{TaghinejadEtAl2021OE}. The heat source was a square wave with a duration of 2\,$\mu$s and was calculated based on a 19\,V pulse height, 30\,Ohm ITO resistance, and $150\,\mu$m$\times$$50\,\mu$m$\times$120\,nm ITO volume.

% \section{Experimental Section}
% \threesubsection{First part of experimental section}\\
% \threesubsection{Second part of experimental section}\\

% \medskip
% \textbf{Supporting Information} \par %Please delete the Suppporting Information statement if it is not applicable. Please supply Supporting Information in another file. Supporting information should not be provided in .tex format
% Supporting Information is available from the Wiley Online Library or from the author.

% Acknowledgements
\medskip
\textbf{Author Contributions} \par R.\,C. performed all experiments and calculations under the supervision of W.\,G. C.\,Y. helped with calculations.

\medskip
\textbf{Acknowledgements} \par %delete if not applicable))
R.\,C., C.\,Y., and W.\,G. acknowledge the support from the National Science Foundation through Grants No. 2316627 and 2428520. 

\medskip
\textbf{Data Availability} \par The data that support the findings of this study are available from the corresponding author upon reasonable request. 
    
\medskip
\textbf{Competing Interests} \par The authors declare that they have no competing financial interests.

% References
\medskip
\bibliographystyle{MSP}
\bibliography{weilu.bib} %% 

\newpage
\begin{figure}
\textbf{Table of Contents}\\
\medskip
  \includegraphics[height=55mm]{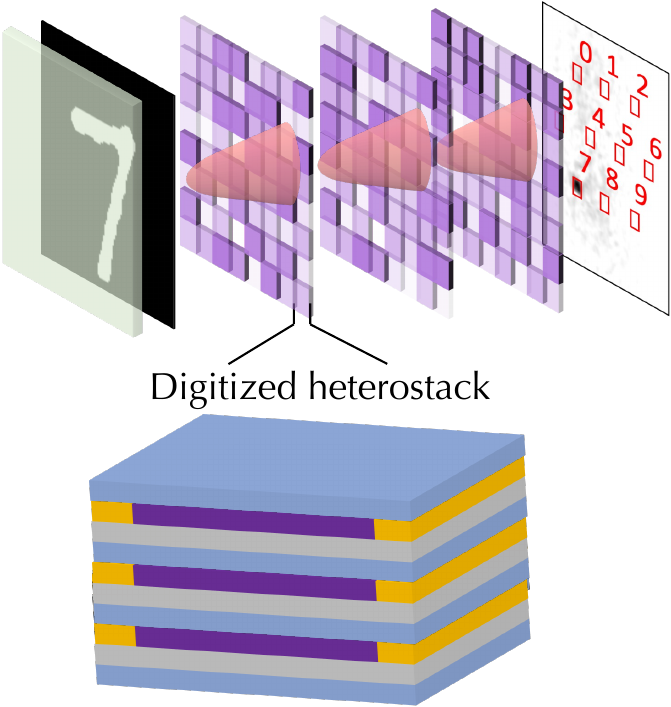}
  \medskip
  \caption*{A phase-change-material-based digitized heterostack is experimentally demonstrated and theoretically analyzed for future energy-efficient, fast reconfigured, and compact diffractive optical neural network systems.}
\end{figure}

\end{document}